\documentclass[12pt]{iopart}
\usepackage{iopams}
\usepackage{graphicx}
\usepackage{cite}
\usepackage{color}
\usepackage[breakwords]{truncate}
\usepackage[font={small}]{caption}
\usepackage{physics}
\usepackage{subcaption}
\usepackage{etoolbox}
\usepackage{verbatim}
\usepackage[section]{placeins}
\usepackage[sectionbib]{chapterbib}
\usepackage{multirow}
\usepackage{pbox}
\usepackage{calc}
\usepackage{braket}
\usepackage[extra]{tipa}
\usepackage{csquotes}
\usepackage[breaklinks,colorlinks = true,urlcolor=blue,citecolor=blue]{hyperref}

\usepackage{color}

\usepackage{epsfig}

\begin{document}

\title[Effect of initial conditions on current fluctuations in active particles]
{Effect of initial conditions on current fluctuations in non-interacting active particles}

\author{Stephy Jose$^1$, Alberto Rosso$^2$ and Kabir Ramola$^3$}

\address{$^{1, 3}$ Tata Institute of Fundamental Research, Hyderabad, India, 500046\\
$^2$ LPTMS, CNRS, Univ.~Paris-Sud, Universit\'e Paris-Saclay, 91405 Orsay, France 
}
\ead{stephyjose@tifrh.res.in, alberto.rosso@universite-paris-saclay.fr, and\\kramola@tifrh.res.in}


\date{\today}

\begin{abstract}
We investigate the effect of initial conditions on the fluctuations of the integrated density current across the origin ($x=0$) up to a given time $t$ in a one-dimensional system of non-interacting run-and-tumble particles. Each particle has initial probabilities $f^+$ and $f^-$ to move with an initial velocity 
$+v$ and $-v$ respectively, where $v>0$.  We derive exact results for the variance (second cumulant) of the current for quenched and annealed averages over the initial conditions for the magnetization and the density fields associated with the particles. We show that at large times, the variance displays a $\sqrt{t}$ behavior, with a prefactor contingent on the specific density initial conditions used. However, at short times, the variance displays either linear $t$ or quadratic $t^2$ behavior, which depends on the combination of magnetization and density initial conditions, along with the fraction $f^+$ of particles in the positive velocity state at $t=0$. Intriguingly, if $f^+=0$, the variance displays a short time $t^2$ behavior with the same prefactor irrespective of the initial conditions for both fields.
\end{abstract}

\section{Introduction}
Active systems are composed of self-propelling particles that consume energy at the individual level to perform directed motion~\cite{vicsek1995novel,czirok1999collective,tailleur2008statistical,lindner2008diffusion,cavagna2010scale,cates2012diffusive,ramaswamy2010mechanics,romanczuk2010collective,romanczuk2012active,martens2012probability,angelani2014first,evans2018run}. The breaking of detailed balance at the microscopic level allows such systems to exhibit behaviors that are distinct from equilibrium systems such as coherent motion, pattern formation, motility-induced phase separation (MIPS), amongst others~\cite{cates2013active,kourbane2018exact,merrigan2020arrested,lee2013active}. Particularly intriguing is the run and tumble (RTP) motion, employed by certain active particles, such as bacteria, to adeptly navigate their surroundings and explore their environment effectively~\cite{evans2018run,malakar2018steady,mori2020universal,mori2020universalp,angelani2014first,martens2012probability,jose2022active,jose2022first,santra2020run,mori2021condensation,dean2021position}. During the running phase, bacteria move towards favorable conditions, while tumbling allows for random reorientation and exploration of new areas. Over the years, the RTP model has attracted significant attention leading to numerous studies on active particles such as first-passage properties, clustering and phase separation, large deviations, and collective motion~\cite{angelani2014first,malakar2018steady,kourbane2018exact,das2018confined,sevilla2019stationary,caprini2019active,jose2022active,jose2022first,de2021survival}.

In this work, we focus on the role of initial conditions on current fluctuations in non-interacting run and tumble particles in one dimension. To explore the effect of initial conditions on particle systems, two types of initial density profiles can be considered: (1) a deterministic profile where the positions of particles are initially fixed (known as the ``quenched density'' setting), and (2) a random profile that allows for fluctuations in the initial positions (known as the ``annealed density'' setting). Active systems, on the other hand, also introduce an additional degree of freedom in the form of magnetization or polarization~\cite{kourbane2018exact,agranov2021exact,agranov2023macroscopic,agranov2022entropy,jose2023current}, which opens up the possibility of two more types of initial conditions: (3) a deterministic initial magnetization profile where the velocities of particles are fixed initially (termed ``quenched magnetization'' setting), and (4) a random initial profile that allows for fluctuations in the initial velocities of the particles (termed ``annealed magnetization'' setting). 

The quantity of interest is the total number of particles $Q$, passing into the half-infinite line ($x>0$) up to time $t$. There have been many studies on the statistics of $Q$ for different passive systems like non-interacting random walkers, symmetric simple exclusion process (SSEP), and Kipnis-Marchioro-Presutti (KMP) model, amongst others using analytical methods like Bethe ansatz, Green's function methods and macroscopic fluctuation theory (MFT)~\cite{derrida2004current,derrida2009current,derrida2009current2,krapivsky2012fluctuations,mallick2022exact,dandekar2022dynamical,dean2023effusion}. The effect of initial conditions on active systems has been first analyzed systematically in~\cite{banerjee2020current} for non-interacting RTPs in one dimension where the exact expression for the variance of the integrated current $Q$ in the annealed density and annealed magnetization setting was derived. Later, this study was extended to incorporate the effect of annealed and quenched magnetization initial conditions in~\cite{jose2023generalized}, focusing on step initial density and zero initial magnetization conditions. In this study, we derive exact expressions of the variance of the integrated current for general step initial conditions for both the density and magnetization fields. This allows for a comparison between the fluctuations due to the differences in initial conditions for both fields at all times. We consider the case where all particles are uniformly distributed to the left of the origin at time $t=0$. Each particle is associated with initial probabilities $f^+$ and $f^-$ to move with velocities $+v$ and $-v$, respectively, where $v>0$. Current fluctuations in related models of run and tumble particle systems have also been analyzed in \cite{jose2023current,tanmoy2023timedependent}. The main focus of our study is to determine the variance of the current for annealed and quenched averages over the initial conditions involving both the density and magnetization fields.

Through our analytical investigation, we find that the variance of the current exhibits a $\sqrt{t}$ behavior at large times with a prefactor contingent on the specific \textit{density} initial conditions used. This demonstrates that the fluctuations in the initial positions of particles have an everlasting effect on the variance of $Q$, unlike the fluctuations in the initial velocities. However, at short times, the variance shows either a linear $t$ or a quadratic $t^2$ behavior influenced by the combination of initial conditions for both the density and magnetization fields, as well as the fraction $f^+$ of the particles with a positive velocity at $t=0$. Such differences in the short-time behavior of variance have also been observed in previous studies of a single active particle. For example, it was shown in Ref.~\cite{basu2018active} that for asymmetric initial magnetization conditions, the mean squared displacement of a single active Brownian particle in two dimensions exhibits non-diffusive behavior at short times and the growth exponent of the variance depends crucially on the initial magnetization initial conditions. A particularly intriguing observation of our study is that when $f^+=0$, meaning there are no particles initially moving with a positive velocity, the variance displays a $t^2$ behavior at short times. Interestingly, this behavior remains the same regardless of the initial conditions of the density and magnetization fields. Another intriguing aspect is the strictly quenched scenario where both the density and magnetization fields are quenched. In this case, the variance always exhibits a $t^2$ behavior at short times regardless of the values of $f^+$ and $f^-$. The specific scenario in which $f^+=f^-=1/2$ has been systematically examined in Ref.~\cite{jose2023generalized}, which is also the usual case studied in the literature, albeit with annealed magnetization initial conditions. In this work, we extend the scope of these results to encompass arbitrary values of both $f^+$ and $f^-$ and focus on the actual quantitative dependence of the fluctuations on the initial conditions in both fields. We argue that the effect of initial conditions goes far beyond the
zero magnetization case studied in Ref.~\cite{jose2023generalized} by considering general step initial conditions for the magnetization field.

This paper is organized as follows. In Sec.~\ref{sec:model}, we introduce the microscopic model and different averages used in the study. In Sec.~\ref{sec:summary}, we provide a summary of the main results. We present derivations of the single particle propagators for different initial bias velocities in Sec.~\ref{sec:propagators}. In Sec.~\ref{sec:fluctuations}, we analytically compute the expressions of the variance of $Q$ for different annealed and quenched initial conditions involving the density and magnetization fields. We present our conclusions from the study in Sec.~\ref{sec:conclusion}. Finally, we provide details pertaining to some of the calculations in~\ref{appendix_laplace_transforms_sq}-\ref{details_calc}.

\begin{figure*}[t!]
\centering
 \includegraphics[width=0.7\linewidth]{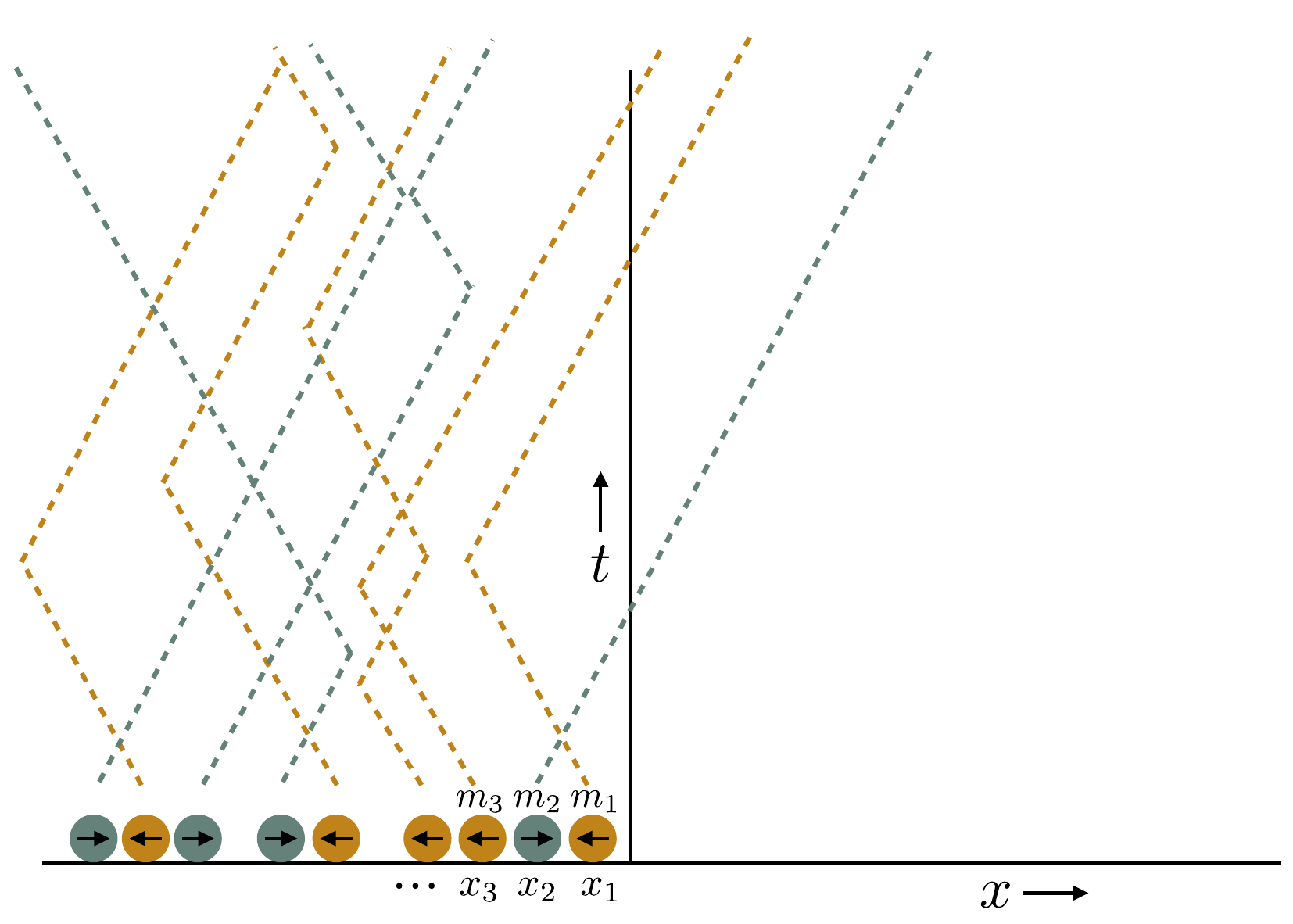}
\caption{Schematic representation of the trajectories of non-interacting run and tumble particles in one dimension. At time $t=0$, all particles are uniformly distributed towards the left of the origin. The label $x_i({m_i})$ indicates the initial position(velocity) of the $i^{\text{th}}$ particle. Different colors denote particles with different initial bias velocities.}
\label{dynamics}
\end{figure*}
\section{Microscopic model}
\label{sec:model}
We consider a one-dimensional box bounded between $\left[-L,0 \right]$ with $N$ run and tumble particles. The dynamics of each particle evolve according to the Langevin equation 
\begin{equation}
\frac{\partial x_i(t)}{\partial t} = v m_i(t),~v>0,~1<i<N.
\label{langevin}
\end{equation} 
The stochastic variable $m_i(t)$ switches values between $+1$ and $-1$ at a fixed rate $\gamma$. For $m_i(t)=\pm 1$, the $i ^{\text{th}}$ particle has a bias velocity $\pm v$ at time $t$. If the initial velocity of a particle is $+v$, then the particle is in $+$ state, and if the initial velocity is $-v$, then the particle is in $-$ state. This velocity behaves akin to an internal spin state, leading us to construct a magnetization field corresponding to the active motion of the particles. We define a density field $\rho(x,t) = L^{-1}\sum_{i=1}^{N} \delta(x - x_i(t))$ associated with the positions and a magnetization field $m(x,t) = L^{-1}\sum_{i=1}^{N} m_i(t) \delta(x - x_i(t))$ associated with the velocities of these active particles. We consider a step initial density profile where all particles are uniformly distributed to the left of the origin with density $\rho$, i.e. $\rho(x,0)=\rho \Theta(-x)$ where $\rho=N/L$ and $\Theta$ is the Heaviside theta function. Let $f^+$ denote the fraction of particles with $+v$ velocity and $f^-$ denote the fraction of particles with $-v$ velocity at time $t=0$, with $f^++f^-=1$. This corresponds to a step initial magnetization profile, $m(x,0)=(f^+-f^-)\rho \Theta(-x)$. For $f^+=f^-=1/2$, $m(x,0)=0$, and this model reduces to the model studied in~\cite{banerjee2020current,jose2023generalized}. Even though we start with a finite-dimensional box, we eventually take the limit $N \rightarrow \infty,~L \rightarrow \infty$ with $N/L \rightarrow \rho$, fixed in our analytical calculations.

We study the statistics of the number of particles $Q$, that cross the origin up to time $t$. As a particle traverses the origin from left to right or vice versa, it adds $+1$ or $-1$ to the current, respectively. Therefore, the integrated current up to time $t$ is exactly equal to the number of particles on the right side of the origin ($x>0$) at time $t$. We provide a schematic representation of the dynamics of the particles in Fig.~\ref{dynamics}. We next elucidate the various types of averages that can be utilized to examine how the fluctuations of $Q$ are influenced by the initial conditions.

\subsection{Annealed and quenched averages}
We consider various types of averages involving the initial positions and the velocities of particles to study the effect of initial conditions on the fluctuations of $Q$. Annealed density (magnetization) setting allows for fluctuations in the initial positions (velocities) of the particles. However, in the quenched density (magnetization) setting, the initial positions (velocities) are fixed. We denote the initial positions and bias states of particles by $\{x_i\}$ and $\{m_i \}$ respectively.
The angular bracket $\langle \cdots \rangle_{\{x_i\},\{m_i\}}$ denotes an average over the history, but with fixed initial positions and velocities of the particles. We also use two additional averages, denoted by $\overline{\cdots}$ for an average over initial positions and $\overbrace{\cdots}$ for an average over initial bias states.

Let us denote by $P_{a,a}(Q,t)$ the probability distribution of $Q$ where both the initial positions and 
velocities are allowed to fluctuate. The initial conditions for positions and velocities are denoted by the first and second subscripts respectively where ``$a$" stands for annealed and ``$q$" stands for quenched scenarios. For this annealed density and annealed magnetization setting, the moment-generating function can be computed as 
\begin{eqnarray}
&&\sum_{Q=0}^\infty e^{-p Q}  P_{a,a}(Q,t)  = \overbrace{ \overline{\langle e^{-p Q}\rangle_{\{x_i\},\{m_i\}}}} .
\label{a_a_av}
\end{eqnarray}
We next consider the case where the initial positions of the particles are allowed to fluctuate, but the velocities are fixed. The flux distribution for this case is represented as $P_{a,q}(Q,t)$. The corresponding moment-generating function is expressed as
\begin{eqnarray}
&&\sum_{Q=0}^\infty e^{-p Q}  P_{a,q}(Q,t)  = \exp \left[ \overbrace{\ln \overline{ \langle e^{-p Q}\rangle_{\{x_i\},\{m_i\}}}}\right] .
\label{a_q_av}
\end{eqnarray}
For the case where both the initial positions and velocities of particles are fixed, the flux distribution is denoted as $P_{q,q}(Q,t)$. The moment-generating function for this flux distribution is given as
\begin{eqnarray}
&&\sum_{Q=0}^\infty e^{-p Q}  P_{q,q}(Q,t)  = \exp \left[\overbrace{ \overline{\ln \langle e^{-p Q}\rangle_{\{x_i\},\{m_i\}}}}\right] .
\label{q_q_qv}
\end{eqnarray}
Finally, we consider the case where the initial positions of the particles are fixed, but the velocities are allowed to fluctuate. The flux distribution for this case is denoted as $P_{q,a}(Q,t)$. The expression for the moment-generating function associated with this process is given as 
\begin{eqnarray}
&&\sum_{Q=0}^\infty e^{-p Q}  P_{q,a}(Q,t)  = \exp \left[ \overline{\ln \overbrace{ \langle e^{-p Q}\rangle_{\{x_i\},\{m_i\}}}}\right] .
\label{q_a_av}
\end{eqnarray}

\section{Summary of the main results}
\label{sec:summary}
\begin{table}[t!]
\centering
\begin{tabular}{ccc}
$t \rightarrow 0$ & $t \rightarrow \infty$ &          \\
\hline
 $\rho v f^+ t-\frac{\rho v \gamma}{4}\left( 3f^+-f^-\right)t^2$   & $\rho \sqrt{\frac{{D_{\text{eff}}}~{t}}{{\pi}}}$   & $\sigma_{a,a}^2(t)$ \\ 
 \hline
  $\rho v f^+ t-\frac{\rho v \gamma}{4}\left( 3f^+-f^-\right)t^2$   & $\rho \sqrt{\frac{{D_{\text{eff}}}~{t}}{{\pi}}}$   & $\sigma_{a,q}^2(t)$ \\  
   \hline
  $\frac{\rho  v \gamma }{4}\left(3 f^++f^- \right)
   t^2$   & $\rho \sqrt{\frac{{D_{\text{eff}}}~{t}}{{2\pi}}}$   & $\sigma_{q,q}^2(t)$ \\ 
      \hline
  $\rho v{f^+f^-}t+\frac{ \rho  v \gamma }{4} \left(3 f^+-f^-\right) \left(f^+-f^-\right) t^2$  & $\rho \sqrt{\frac{{D_{\text{eff}}}~{t}}{{2\pi}}}$ & $\sigma_{q,a}^2(t)$ \\ 
\end{tabular}
 \caption{Asymptotic behaviors of the variance $\sigma^2(t)$ of the integrated current for different initial conditions. The initial conditions for density and magnetization fields are denoted by the first and second subscripts respectively where ``$a$" stands for annealed and ``$q$" stands for quenched. Here, $D_{\text{eff}}=v^2/(2\gamma)$ is the effective diffusion constant for RTP motion in one dimension. The mean of $Q$ denoted as $\mu(t)$ is equal to $\sigma_{a,a}^2(t)=\sigma_{a,q}^2(t)$ and is the same for all initial conditions when the initial positions are randomized.}
\label{table1}
\end{table}

We focus on the role of initial conditions on the fluctuations of the integrated current $Q$. The mean of $Q$ is a self averaging property, i.e.~it takes on the same characteristics regardless of the specific initial conditions~\cite{krapivsky2012fluctuations}. However, the variance displays interesting differences for various averages involving the initial conditions. We first consider the annealed density and annealed magnetization setting which allows for fluctuations in both the positions and velocities of particles at time $t=0$. Another equivalent scenario is the annealed density and quenched magnetization setting where the positions are allowed to fluctuate, but the velocities are fixed at time $t=0$~\cite{jose2023generalized}. In the infinite system size limit, the distributions of $Q$ for both these initial conditions become identical. This is due to the fact that whether the velocities are kept constant or allowed to fluctuate makes no difference as the initial positions of particles are randomized. Therefore, the fluctuations in the initial velocities do not have any effect on the distribution of $Q$ (and hence the variance) when the initial positions are randomized. 
The exact expression for the variance of $Q$ for these cases can be computed as 
\begin{equation}
 \sigma_{a,a}^2(t) = \sigma_{a,q}^2(t)=\frac{\rho  v}{4 \gamma } \left[2 e^{-t \gamma } t \gamma   \left(\pmb{I}_0(t \gamma )+\pmb{I}_1(t
   \gamma )\right)+\left(f^+-f^-\right) \left(1-e^{-2 t \gamma }\right)\right].
   \label{mu_t_exact}
\end{equation}
As before, the initial conditions for density and magnetization fields are denoted by the first and second subscripts respectively where ``$a$" stands for annealed and ``$q$" stands for quenched scenarios. The symbols $\pmb{I}_0$ and $\pmb{I}_1$ denote modified Bessel functions of order $0$ and $1$ respectively. The modified Bessel function $\pmb{I}_n(z)$ of order $n$ is a solution to the homogeneous Bessel differential equation $z^2y^{''}(z)+z y^{'}(z)-(z^2+n^2)y=0$.
The expression in Eq.~\eqref{mu_t_exact} is also equal to the mean for all initial conditions. For $f^+=f^-=1/2$, we obtain the simplified expression~\cite{banerjee2020current,jose2023generalized}
\begin{eqnarray}
\sigma_{a,a}^2(t) = \sigma_{a,q}^2(t)=    \frac{ \rho v}{2} t e^{-t \gamma } (\pmb{I}_0(t \gamma )+\pmb{I}_1(t \gamma )) ,~\text{for}~ f^+=f^-=1/2.
\label{mu_t_sym}
\end{eqnarray}
This is the specific case studied in~\cite{banerjee2020current,jose2023generalized}. Unlike the symmetric case, fluctuations in the asymmetric case involve an additional exponential term as illustrated in Eq.~\eqref{mu_t_exact}. Interestingly, this term becomes exactly equal to zero in the symmetric case. This demonstrates the effect of asymmetry in the initial magnetization initial conditions explicitly.
\begin{figure*}[t!]
\centering
 \includegraphics[width=1.0\linewidth]{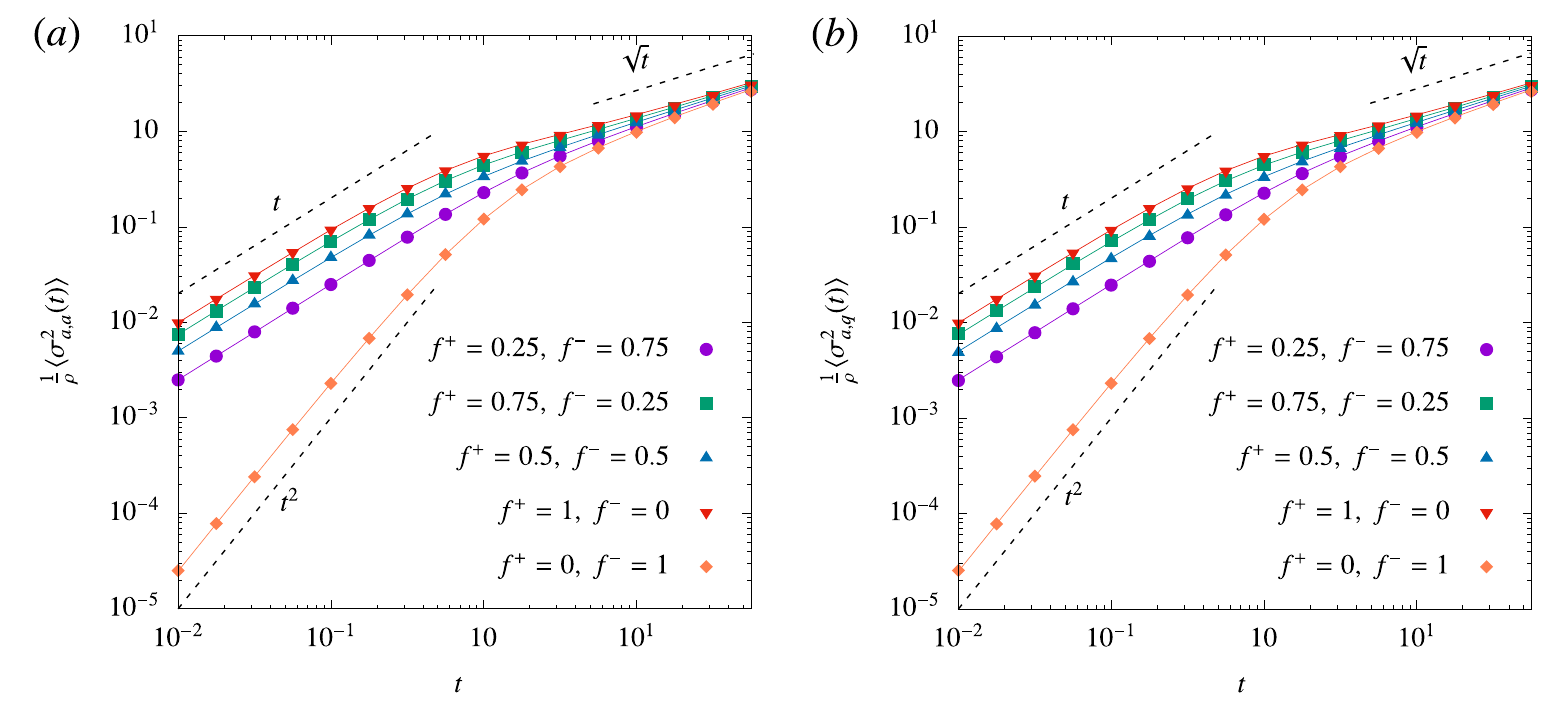}
\caption{Variance of the time integrated current $Q$ plotted as a function of time $t$ for different values of $f^+$ and $f^-$. The specific cases displayed are (a)~annealed density and annealed magnetization initial conditions~(b)~annealed density and quenched magnetization initial conditions. The points are obtained from direct numerical simulations and the solid curves correspond to the exact analytical result in Eq.~\eqref{mu_t_exact}. These plots are for the parameter values $\rho = 20,~ \gamma = 1,~ v= 1$. We notice that the variance exhibits similar behavior for both annealed and quenched magnetization initial conditions. Thus we infer that magnetization initial conditions do not influence the statistics of $Q$ when the density initial conditions are annealed, regardless of the value of $f^+$. We also observe that the case where $f^+=0$ is distinct from the others, displaying a $t^2$ behavior at short times.}  
\label{fig_annealed_density}
\end{figure*}

The third case we study is the quenched density and quenched magnetization setting where both the positions and velocities of particles are fixed initially. For this non-trivial case, the exact expression for the variance in real time is hard to compute. Nevertheless, it is possible to compute the exact expression of the variance in Laplace space. We define the Laplace transform of a function $f(t)$ as $\mathcal{L}\left[ f(t)\right]=\tilde f (s)=\int_0^\infty dt e^{-s t} f(t)$.  We show that
\begin{equation}
\tilde \sigma_{q,q}^2 (s) =\rho \left[(f^+-f^-) \left(\frac{v}{4 s (s+2 \gamma )}-\frac{v
   K\left(-\frac{8 \gamma  (s+2 \gamma )}{s^2}\right)}{2 \pi  s (s+2 \gamma )}\right)+\frac{v \gamma }{s (s+2 \gamma ) \sqrt{s (s+4 \gamma )}} \right].
\label{sigma_q_q}
\end{equation}
In the above equation, $K$ is the elliptic integral of the first kind defined as 
\begin{equation}
K(m)=\int_{0}^{\frac{\pi}{2}} d\theta~{1}/{\sqrt{1-m~{\sin}^2\theta }}.
\end{equation}
Equation~\eqref{sigma_q_q} can be used to extract the asymptotic behaviors of the variance in real-time and is given in Table~\ref{table1}. Additionally, for symmetric initial conditions with $f^+=f^-=1/2$, this expression can be inverted exactly yielding~\cite{jose2023generalized}
\begin{equation}
\sigma_{q,q}^2 (t) = \frac{\rho v}{4} t  e^{-2 \gamma  t} \left[(2+\pi  \pmb{L}_0(2 t \gamma )) \pmb{I}_1(2 t
 \gamma )- \pi \pmb{L}_1(2 t \gamma ) \pmb{I}_0(2 t \gamma )\right],~\text{for}~f^+=f^-=1/2,
 \label{sigma_q_q_sym}
\end{equation}
where $\pmb{L}_0$ and $\pmb{L}_1$ are modified Struve functions of order $0$ and $1$ respectively. The modified Struve function $\pmb{L}_n(z)$ of order $n$ is a solution to the non-homogeneous Bessel differential equation $z^2y^{''}(z)+z y^{'}(z)-(z^2+n^2)y={4{\left( {z}/{2} \right)}^{n+1}}/{\left(\sqrt{\pi}\Gamma(n+\frac{1}{2})\right)}$.

The final case we analyze is the quenched density and annealed magnetization setting where the positions of particles are fixed, but the velocities are allowed to fluctuate at time $t=0$.
 For these initial conditions,
the exact expression for the variance in Laplace space can be obtained as
\begin{eqnarray}
\tilde \sigma_{q,a}^2 (s) &=&\frac{\rho  v}{s^{3/2} (s+2 \gamma )}\Bigg [ \left(\frac{\gamma }{\sqrt{s+4 \gamma }}-\frac{1}{2}
   \sqrt{s+2 \gamma }\right)
   \left({f^+}^2+{f^-}^2\right
   )\nonumber\\
&-&\left(\frac{2 \gamma }{\sqrt{s+4 \gamma
   }}+\sqrt{s+2 \gamma }-\sqrt{s+4 \gamma }\right)
   f^+ f^- +\frac{1}{2} \sqrt{s+2 \gamma }\nonumber\\
   &-&\sqrt{s} \left(\frac{K\left(-\frac{8 \gamma  (s+2
   \gamma )}{s^2}\right)}{2 \pi }+\frac{1}{4}\right)
   \left({f^+}^2-{f^-}^2\right
   )+\frac{1}{2} \sqrt{s}
   \left(f^+-f^-\right)
   \Bigg].\nonumber\\
\label{sigma_q_a}
\end{eqnarray}
For symmetric initial conditions $f^+=f^-=1/2$, the above expression can be inverted exactly yielding~\cite{jose2023generalized} 
\begin{equation}
\sigma_{q,a}^2 (t) = \frac{\rho v}{8} t  e^{-2 \gamma  t} \left[(4+\pi  \pmb{L}_0(2 t \gamma )) \pmb{I}_1(2 t
 \gamma )+(2-\pi  \pmb{L}_1(2 t \gamma )) \pmb{I}_0(2 t \gamma )\right],~\text{for}~f^+=f^-=1/2.
 \label{sigma_q_a_sym}
\end{equation} 
\begin{figure*}[t!]
\centering
 \includegraphics[width=1.0\linewidth]{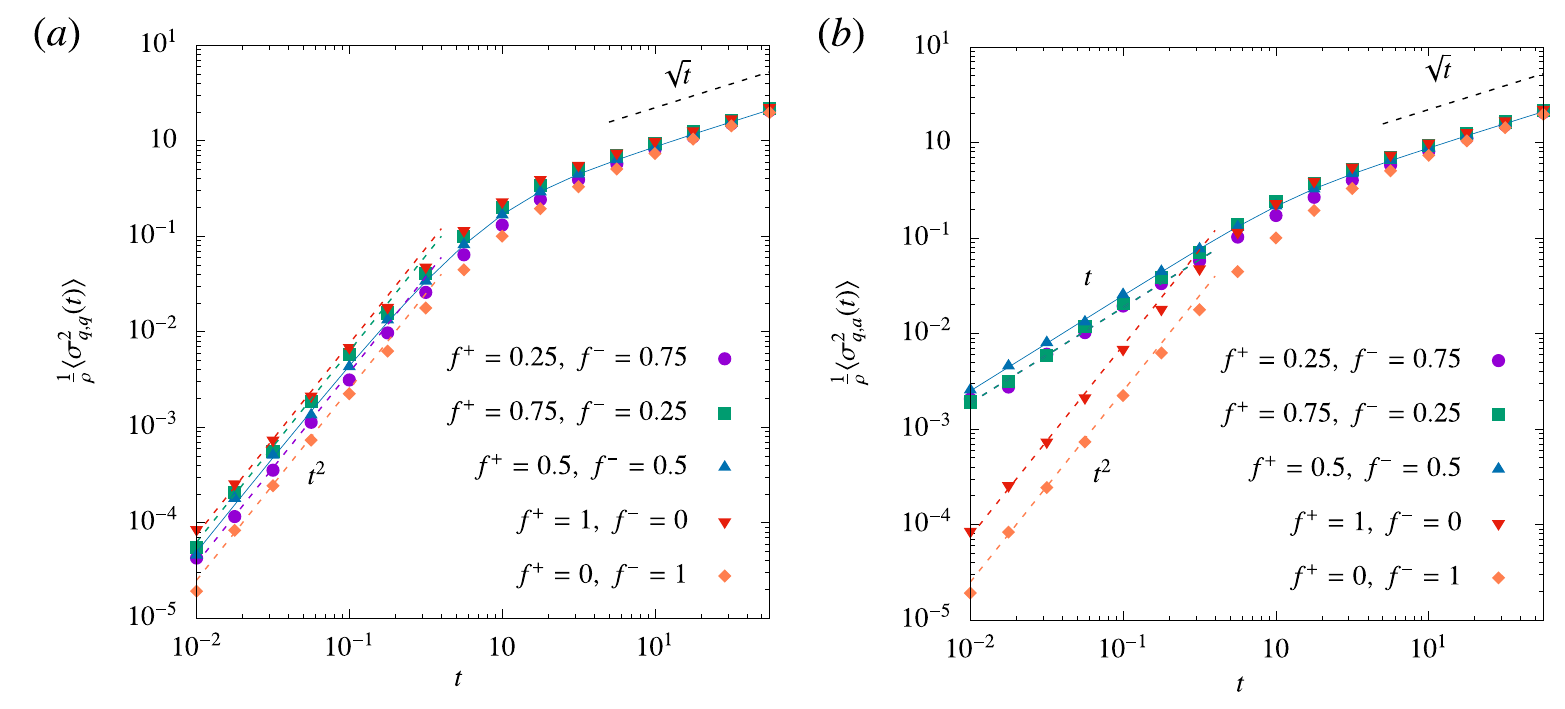}
\caption{Variance of the time integrated current $Q$ plotted as a function of time $t$ for different values of $f^+$ and $f^-$. The specific cases displayed are (a)~quenched density and quenched magnetization initial conditions~(b)~quenched density and annealed magnetization initial conditions. The points are obtained from direct numerical simulations. The solid curves for $f^+=f^-=0.5$ in (a) and (b) correspond to the exact analytical results in Eqs.~\eqref{sigma_q_q_sym}~and~\eqref{sigma_q_a_sym} respectively. The dashed curves correspond to the asymptotic behaviors listed in Table~\ref{table1}. These plots are for the parameter values $\rho = 20,~ \gamma = 1,~ v= 1$. We observe that quenched density and quenched magnetization initial conditions consistently result in suppressed fluctuations at short times, regardless of the value of $f^+$. We also notice that quenched density and annealed magnetization initial conditions differ from the other cases, particularly in the behavior observed for $f^+=0$ and $f^+=1$ cases.}
\label{fig_quenched_density}
\end{figure*}
The asymptotic behaviors of the variance for each of the four cases discussed above are listed in Table~\ref{table1}. We also compare our analytical predictions for the variance of $Q$ with direct Monte Carlo simulations in Fig.~\ref{fig_annealed_density}~and~Fig.~\ref{fig_quenched_density}. Our theoretical predictions align remarkably well with the results obtained from Monte Carlo simulations.

We notice from Table~\ref{table1} that for the special case where $f^+=0$ and $f^-=1$, the mean and the variance always grow as ${\rho v \gamma t^2}/{4}$ in the small time limit irrespective of the initial conditions. This results in suppressed fluctuations, and the short-time behavior of current fluctuations becomes independent of the initial conditions. This independence arises because, in the absence of particles initially in the positive state, the sole factor contributing to the current across the origin is the flipping of particle states. Consequently, within this flipping time scale, all initial conditions essentially become identical. The behavior of the mean can be understood as explained below. Consider a single RTP starting its motion from the location $-x_0$ with $x_0>0$ at time $t=0$ in $-$ state (i.e. with velocity $-v$). The particle cannot cross the origin until it flips the velocity state. At short times, we can safely approximate that the mean is dominated by trajectories with a single flip. Let $\tau$ be the time taken by the particle to flip its velocity state. Here, $\tau$ is a stochastic variable. For the particle to be able to cross the origin within a time $t>\tau$, the necessary condition is
\begin{equation}
    x_0+\tau v < (t-\tau) v.
\label{inequality_cond}
\end{equation}
That is, the distance to the location of the particle just before the first flip (before time $\tau$) should be less than the distance traveled by the particle in the remaining time interval $t-\tau$. Equation~\eqref{inequality_cond} can be rewritten as 
\begin{equation}
\tau < \frac{1}{2}\left( t-\frac{x_0}{v}\right).
\end{equation}
Since the distribution of the time gap between consecutive flips is Poissonian, the probability that the particle will cross the origin within a time $t$ is given as
\begin{equation}
    \langle Q(t)|x_0\rangle=\int_0^{\frac{1}{2}\left( t-\frac{x_0}{v}\right)}~ d \tau~ \gamma e^{-\gamma \tau}=1-e^{-\frac{\gamma}{2}\left( t-\frac{x_0}{v}\right)}.
\end{equation}
If the density of particles is $\rho$,
the average current is then given as
\begin{equation}
\langle Q \rangle=
    \rho \int_0^{v t}dx_0~\left( 1-e^{-\frac{\gamma}{2}\left( t-\frac{x_0}{v}\right)}\right)=\rho v \left(t-\frac{2 \left(1-e^{-\frac{1}{2} t \gamma
   }\right)}{\gamma }\right),
\end{equation}
which in the short time limit yields
\begin{eqnarray} 
\langle Q \rangle\xrightarrow[t \rightarrow 0]{}\frac{ \rho v \gamma}{4} t^2.
\label{meanfminus1}
\end{eqnarray}
In Sec.~{\ref{sec:fluctuations}}, we demonstrate that when the initial density conditions are annealed, the flux distribution consistently follows a Poisson distribution. This holds true regardless of the initial magnetization conditions. Therefore, both $P_{a,a}(Q,t)$ and $P_{a,q}(Q,t)$ are Poissonian, as evidenced by Eqs.~\eqref{Paa} and~\eqref{Paq}, respectively. Consequently, the mean value of the distribution is equal to its variance in these cases. Hence, the result presented in Eq.~\eqref{meanfminus1} also applies to the variance when the initial density conditions are annealed. This finding aligns with the limiting behavior outlined in Table~\ref{table1}, considering the parameter choice of $f^+=0$ and $f^-=1$. We also note that for initial conditions where both the density and magnetization fields are quenched, the variance consistently follows a $t^2$ behavior at short times, regardless of the specific values of $f^+$ and $f^-$.



\section{Single particle propagators}
\label{sec:propagators}

In this section, we provide expressions for the single particle propagators associated with a run and tumble particle (RTP) in one dimension. We also provide expressions for the integrals of these Green's functions. These expressions will be useful in the analytical calculations presented in the next section. We consider symmetric and asymmetric initial bias velocities separately. An alternate method to derive the single particle propagators has also been given in~\cite{jose2023generalized}. Here, we employ a Laplace transform approach to the Fokker-Planck equations governing occupation probabilities. We then solve these equations with initial conditions and normalization constraints. In contrast, the derivation outlined in~\cite{jose2023generalized} involves solving the matrix equation associated with the Fourier-Laplace transform of the occupation probabilities.

We consider an RTP starting its motion from the position $x=0$ at time $t=0$ in one dimension. We use the notation $P_m(x,t)$ to represent the probability density of the particle to be at location $x$ at time $t$ in the velocity state $m$, where $m=\pm $. The evolution equations for this probability density are ~\cite{malakar2018steady}
\begin{eqnarray}
\frac{\partial P_+(x,t)}{\partial t}
&=& -v \frac{\partial P_+(x,t)}{\partial x} - \gamma P_+(x,t) + \gamma P_-(x,t),  \nonumber\\
\frac{\partial P_-(x,t)}{\partial t}
&=& +v \frac{\partial P_-(x,t)}{\partial x} - \gamma P_-(x,t) + \gamma P_+(x,t).  \label{Pequations}
\end{eqnarray}
The total probability density of being at position $x$ at time $t$ is given as 
\begin{equation}
    P(x,t)=P_+(x,t)+P_-(x,t).
\end{equation}
Taking a Laplace transform of Eq.~\eqref{Pequations} yields
\begin{eqnarray}
-P_+(x,0)
+ v \frac{\partial \tilde P_+(x,s)}{\partial x} +(s+ \gamma) \tilde P_+(x,s) - 
\gamma \tilde P_-(x,s) &=&0,  
\nonumber\\
-P_-(x,0)
- v \frac{\partial \tilde P_-(x,s)}{\partial x} +(s+ \gamma) \tilde P_-(x,s) - 
\gamma \tilde P_+(x,s) &=&0  \label{Peqations_s} .
\end{eqnarray}
 We consider an ansatz of the form
\begin{eqnarray}
\tilde P_\pm= A_\pm {\rm e^{-\lambda x}}  \quad\mbox{for}\quad x>0,\nonumber\\
\tilde P_\pm = B_\pm {\rm e^{ +\lambda x}}  \quad\mbox{for}\quad x<0.
\label{ansatz}
\end{eqnarray}
Substituting these solutions into Eq.~\eqref{Peqations_s} (away from $x=0$) yields 
\begin{eqnarray}
(s+\gamma -\lambda v) A_+ =\gamma A_-, \nonumber\\
(s+\gamma +\lambda v) B_+ =\gamma B_-, \nonumber\\
(s+\gamma +\lambda v) A_- =\gamma A_+,
\nonumber\\
(s+\gamma -\lambda v) B_- =\gamma B_+.\label{cond} 
\end{eqnarray}
Combining the above equations yields the expression for $\lambda$ as
\begin{equation}
\lambda =  \frac{\sqrt{s(s+2\gamma)}}{v}.
\label{lambda}
\end{equation}
Since the total probability density $P(x,t)=P_+(x,t)+P_-(x,t)$ has to be normalised, we obtain the condition
\begin{equation}
\int_{- \infty}^{\infty} d x \left( \tilde P_+ + \tilde P_-\right) = \frac{1}{s}.
\end{equation}
This implies
\begin{equation}
A_+ + A_- + B_+ + B_- = \frac{\lambda}{s}.
\label{normalisation}
\end{equation}
This condition along with the initial conditions can be together used to evaluate the undetermined coefficients $A_{\pm}$ and $B_{\pm}$
as we demonstrate in the following subsections. We consider symmetric and asymmetric initial bias velocities separately.

\subsection{Symmetric initial bias velocity}
We consider symmetric initial bias velocities of the form
\begin{equation}
P_+(x,0) = P_-(x,0) = \frac{1}{2} \delta(x),
\label{sym_cond}
\end{equation}
where the particle starts from $+$ or $-$ velocity state with equal probability. Integrating Eq.~\eqref{Peqations_s} over $x$ yields
\begin{eqnarray}
-\frac{1}{2}+\frac{(s+\gamma)}{\lambda}(A_++B_+)-\frac{\gamma}{\lambda}(A_-+B_-)=0, 
\nonumber\\ 
-\frac{1}{2}+\frac{(s+\gamma)}{\lambda}(A_-+B_-)-\frac{\gamma}{\lambda}(A_++B_+)=0.
\label{Peq_s2} 
\end{eqnarray}
Solving Eqs.~\eqref{cond},~\eqref{normalisation}~and~\eqref{Peq_s2} yields the expressions for coefficients as
\begin{eqnarray}
A_+ = \frac{v \lambda+s }{4 s v},
&\quad& A_- = \frac{v \lambda -s}{4 s v},\nonumber\\
B_+ = \frac{v \lambda -s}{4 s v},
&\quad& B_- = \frac{v \lambda+s }{4 s v}.
\end{eqnarray}
Using the above expressions and the form of ansatz provided in Eq.~\eqref{ansatz}, we directly obtain~\cite{banerjee2020current}
\begin{equation}
\tilde P(x,s) = \tilde P_+(x,s)+ \tilde P_-(x,s)
=  {\rm e}^{-|x|\lambda}\frac{\lambda }{2s},
\label{ptot_sym}
\end{equation}
where the expression for $\lambda$ is provided in Eq.~\eqref{lambda}. 

We define the symmetric Green's function $G^0(x,x_i,t)$ as the probability density of finding an RTP at position $x$ at time $t$, given that it started from position $x_i$ at time $t = 0$, with an equal chance of being in the $+$ or $-$ state. The superscript $``0"$ indicates the symmetric scenario where the particle has an equal probability of being in the $+$ or $-$ state at time $t=0$. Since the evolution equations are invariant under translations, Eqs.~\eqref{ptot_sym}~and~\eqref{lambda} directly lead to
\begin{eqnarray}
\tilde G^0(x,-z,s)&=& \frac{e^{-\frac{\left|
   x+z\right|  \sqrt{s (s+2
   \gamma )}}{v}} \sqrt{s (s+2
   \gamma )}}{2 v s},\quad z=-x_i.
\label{greens_functions_laplace}
\end{eqnarray}
Here, $G^0(x,-z,s)$ represents the Laplace transform of the Green's function for an RTP that starts with equal probabilities to be in both the $+$ and $-$ states. For step initial conditions, we consider $x_i\le0$, thus $z\ge 0$.

We define the integral of the Green's function $G^0(x,-z,t)$ over the half-infinite line as
\begin{eqnarray}
U^0(z,t) = \int_0^\infty dx~G^0(x,-z,t), \quad z\geq 0 .
\label{uzt}
\end{eqnarray} 
The quantity $U^{0}(z,t)$ corresponds to the probability that a particle starting from the location $-z$ in a symmetrized velocity state is found towards the right side of the origin at time $t$. 
Using the exact expression in Eq.~\eqref{greens_functions_laplace}, we derive the expression for $U^0(z, t)$ in Laplace space as 
\begin{eqnarray}
\tilde U^0(z,s) &=& \frac{\exp\left(-z\frac{\sqrt{s(s+2\gamma)}}{v} \right)}{2s}.
\label{uzs}
\end{eqnarray}
The integral of the function $\tilde U^0(z,s)$ over $z$ yields
\begin{eqnarray}
\int_0^\infty dz~\tilde U^0(z,s) =\frac{v}{2 s^{3/2}\sqrt{(s+2 \gamma)}}. 
\label{Us_int}
\end{eqnarray}
We next define the Laplace transform of the square of the function $U^0(z,t)$ as
\begin{equation}
\tilde V^0(z,s)=\mathcal{L}\left[{U^0(z,t)}^2\right].
\label{Vs_def}
\end{equation}
This is a useful quantity that enters the computations of the current fluctuations presented in the next section.
The integral of the function $\tilde V^0(z,s)$ over $z$ yields 
\begin{eqnarray}
\int_0^\infty dz~\tilde V^0(z,s) =\frac{ v}{2 s^{3/2}} \left(\frac{1}{\sqrt{s+2 \gamma }}-\frac{\sqrt{s+4 \gamma }}{2( s+2\gamma)}\right). 
\label{Vs_int}
\end{eqnarray}
We provide details regarding the calculation of the above integral in~\ref{details_calc}.

\subsection{Asymmetric initial bias velocity}
We next consider asymmetric initial bias velocities where the particle is initialized in either $+$ or $-$ state.\\\\
Case 1: Particle initialized in $+$ state:\\\\
We consider asymmetric initial conditions of the form
\begin{equation}
P_+(x,0) = \delta(x),\quad P_-(x,0) = 0.
\label{asym_cond_plus}
\end{equation}
Here, the particle starts from $+$ velocity state at time $t=0$. Integrating Eq.~\eqref{Peqations_s} over $x$ yields
\begin{eqnarray}
-1+\frac{(s+\gamma)}{\lambda}(A_++B_+)-\frac{\gamma}{\lambda}(A_-+B_-)&=&0, 
\nonumber \\
\frac{(s+\gamma)}{\lambda}(A_-+B_-)-\frac{\gamma}{\lambda}(A_++B_+)&=&0.
\label{Peqs_s3} 
\end{eqnarray}
Solving Eqs.~\eqref{cond},~\eqref{normalisation}~and~\eqref{Peqs_s3} yields the expressions for coefficients as
\begin{eqnarray}
A_+ = \frac{s+\gamma+v \lambda}{2 v^2 \lambda},
&\quad& A_- = \frac{\gamma}{2 v^2 \lambda},\nonumber\\
B_+ = \frac{s+\gamma-v \lambda}{2 v^2 \lambda},
&\quad& B_- = \frac{\gamma}{2 v^2 \lambda}.
\end{eqnarray}
Using the above expressions and the form of ansatz provided in Eq.~\eqref{ansatz}, we obtain
\begin{equation}
\tilde P(x,s) = \tilde P_+(x,s)+ \tilde P_-(x,s)
=  \frac{{\rm e}^{- |x|\lambda}}{2s}\left(\lambda+\text{sgn}(x)\frac{s}{v}\right),
\label{ptot_asym_plus}
\end{equation}
where the expression for $\lambda$ is provided in Eq.~\eqref{lambda} and \text{sgn} is the sign function.
\\\\
Case 2: Particle initialized in $-$ state:\\\\
We consider asymmetric initial conditions of the form
\begin{equation}
P_-(x,0) = \delta(x),\quad P_+(x,0) = 0.
\label{asym_cond_minus}
\end{equation}
Here, the particle starts from $-$ velocity state at time $t=0$. Integrating Eq.~\eqref{Peqations_s} over $x$ yields
\begin{eqnarray}
\frac{(s+\gamma)}{\lambda}(A_++B_+)-\frac{\gamma}{\lambda}(A_-+B_-)&=&0, 
\nonumber \\
-1+\frac{(s+\gamma)}{\lambda}(A_-+B_-)-\frac{\gamma}{\lambda}(A_++B_+)&=&0.
\label{Peqs_s4} 
\end{eqnarray}
Solving Eqs.~\eqref{cond},~\eqref{normalisation}~and~\eqref{Peqs_s4} yields the expressions for coefficients as
\begin{eqnarray}
A_+ = \frac{\gamma}{2 v^2 \lambda},
&\quad& A_- =\frac{s+\gamma-v \lambda}{2 v^2 \lambda},\nonumber\\
B_+ = \frac{\gamma}{2 v^2 \lambda},
&\quad& B_- = \frac{s+\gamma+v \lambda}{2 v^2 \lambda}.
\end{eqnarray}
Using the above expressions, we directly obtain
\begin{equation}
\tilde P(x,s) = \tilde P_+(x,s)+ \tilde P_-(x,s)
=  \frac{{\rm e}^{- |x|\lambda}}{2s}\left(\lambda-\text{sgn}(x)\frac{s}{v}\right),
\label{ptot_asym_minus}
\end{equation}
where the expression for $\lambda$ is provided in Eq.~\eqref{lambda}.

We define the Green's functions $G^\pm(x,x_i,t)$, which represent the probability density of finding an RTP at position $x$ at time $t$, given that it started from position $x_i$ at time $t=0$, in a fixed velocity state $\pm$. The superscript $``\pm"$ indicates the asymmetric scenario where the particle starts from either the $+$ or $-$ state at time $t=0$ with a probability $1$. Since the evolution equations are invariant under translations, Eqs.~\eqref{ptot_asym_minus},~\eqref{ptot_asym_plus}~and~\eqref{lambda} directly yield
\begin{equation}
\tilde G^\pm(x,-z,s) = \frac{e^{-\frac{\left|
   x+z\right|  \sqrt{s (s+2
   \gamma )}}{v}}
   \left(\sqrt{s (s+2 \gamma
   )} \pm s~
   \text{sgn}(x+z)\right)}{2 v
   s},\quad z=-x_i,
\label{greens_functions_laplace_plus_minus}
\end{equation}
where \text{sgn} denotes the sign function.

We next define the integral of the Green's function $G^\pm(x,-z,t)$ over the half-infinite line $x \geq 0$ as
\begin{eqnarray}
U^\pm(z,t) = \int_0^\infty dx~ G^\pm(x,-z,t).
\label{uzt_plus_minus}
\end{eqnarray}
The quantity $U^{\pm}(z,t)$ corresponds to the probability that a particle starting from the location $-z$ in the velocity state $\pm$ is found towards the right side of the origin at time $t$.
Using Eq.~\eqref{greens_functions_laplace_plus_minus} and the definition provided in Eq.~\eqref{uzt_plus_minus}, we obtain the exact expression for the Laplace transform of $U^\pm(z,t)$ as
\begin{eqnarray}
\tilde U^\pm(z,s) &=& \frac{e^{-\frac{z \sqrt{s (s+2
   \gamma )}}{v}}}{2
   s}
   \left(1 \pm \frac{s}{\sqrt{s
   (s+2 \gamma )}}\right).
\label{uzs_plus_minus}
\end{eqnarray}
We integrate the function $\tilde U^\pm(z,s)$ over $z$ to yield
\begin{eqnarray}
\int_0^\infty dz~\tilde U^\pm(z,s) =\frac{ v \left(\sqrt{s (s+2
   \gamma )} \pm s\right)}{2 s^2
   (s+2 \gamma )} . 
   \label{Uplus_minus_int}
\end{eqnarray}
We next define 
\begin{equation}
\tilde V^\pm(z,s)=\mathcal{L}\left[{U^\pm(z,t)}^2\right],
\label{Vplus_minus_def}
\end{equation}
as the Laplace transform of the square of the function $U^\pm(z,t)$. After performing some algebraic calculations (details are presented in~\ref{details_calc}), it can be shown that 
\begin{equation}
\int_0^\infty dz~\tilde V^\pm(z,s) =\frac{v }{s (s+2 \gamma )}\left(\pm \frac{1}{4}+\frac{1}{2} \sqrt{\frac{s+2 \gamma }{s}}-\frac{\gamma }{\sqrt{s
   (s+4 \gamma )}} \pm \frac{K\left(-\frac{8 \gamma  (s+2 \gamma )}{s^2}\right)}{2 \pi
   }\right). 
   \label{Vplus_minus_int}
\end{equation}

Another useful quantity that enters the computations of current fluctuations is the Laplace transform of the product of the functions $U^+(z,t)$ and $U^-(z,t)$. We denote
\begin{equation}
\tilde V^{\text{cross}}(z,s)=\mathcal{L}\left[U^+(z,t)U^-(z,t)\right],
\end{equation}
as the Laplace transform of the product of the functions $U^+(z,t)$ and $U^-(z,t)$. 
 As the symmetric Green's function $G^0(x,x_i,t)$ is the average of the asymmetric Green's functions $G^+(x,x_i,t)$ and $G^-(x,x_i,t)$, we have
\begin{equation}
G^0(x,x_i,t)=\left( G^+(x,x_i,t)+G^-(x,x_i,t)\right)/2,
\label{G_0_def}
\end{equation}
and consequently
\begin{equation}
U^0(z,t)=\left( U^+(z,t)+U^-(z,t)\right)/2.
\label{U_0_def}
\end{equation}
Using Eqs.~\eqref{Vs_def},~\eqref{Vplus_minus_def},~and \eqref{U_0_def} we obtain
\begin{equation}
\tilde V^{\text{cross}}(z,s)=2 \tilde V^{0}(z,s)-\frac{1}{2}\left(\tilde V^{+}(z,s)+\tilde V^{-}(z,s) \right).
\end{equation}
After performing the integration over $z$ and plugging in the expressions provided in Eqs.~\eqref{Vs_int} and \eqref{Vplus_minus_int}, we obtain 
\begin{equation}
\int_0^\infty dz~\tilde V^{\text{cross}}(z,s) =\frac{v }{\sqrt{s} (s+2 \gamma )^{3/2}}\left(\frac{1}{2}+\frac{\gamma }{s}+\frac{\gamma  \sqrt{\frac{s (s+2 \gamma )}{s+4
   \gamma }}}{s^{3/2}}-\frac{\sqrt{s (s+2 \gamma ) (s+4 \gamma )}}{2
   s^{3/2}}\right).
    \label{Vcross_int}
\end{equation}

\section{Current fluctuations for different initial conditions}
\label{sec:fluctuations}
In this section, we analytically compute the variance of $Q$ for different initial conditions involving the density and magnetization fields. Let $\{x_i\}$ denote the positions and $\{m_i \}$ denote the velocities or bias states of particles at time $t=0$. Each position $x_i$ is drawn from a uniform distribution between $-L$ and $0$. Here, $i \in [1,N]$ denotes the particle index with $x_i<0$. The initial bias state $m_i$ can be $+$ with probability $f^+$ or $-$ with probability $f^-$, with $f^++f^-=1$. Thus $f^+$ and $f^-$ denote the fraction of positive and negative biased particles respectively at time $t=0$.  

We follow similar steps introduced in~\cite{banerjee2020current,jose2023generalized} to compute the current fluctuations for different initial conditions. Let ${\cal I}_i(t)$ be an indicator function defined as
\begin{equation}
    {\cal I}_i(t)= 
\begin{cases}
   1,&\text{if the $i^{\rm th}$ particle
is towards the right side of the origin at time $t$},\\
     0,&\text{otherwise} .
\end{cases}
\end{equation}
The total number of particles $N^+$ to the right side of the origin at time $t$ is thus given as
\begin{equation}\label{N}
 N^+= \sum_{i=1}^N {\cal I}_i(t) .
\end{equation}
As mentioned before, the number of particles that cross the origin up to time $t$ is equal to the number of the particles on the right side of the origin at time $t$. For a fixed initial realization of the positions $\{x_i\}$ and the bias states $\{m_i\}$, the distribution of $Q$ is given as
\begin{eqnarray}\label{PQ}
P(Q,t,\{ x_i\},\{ m_i\}) = {\rm Prob.}~(N^+ = Q) = \left \langle \delta \left[Q-\sum_{i=1}^N {\cal I}_i(t)\right]\right \rangle_{\{x_i\},\{m_i\}} .  
\end{eqnarray}
Here, the angular bracket $\langle \cdots \rangle_{\{x_i\},\{m_i\}}$ denotes an average over the history, but with fixed initial positions $\{x_i\}$ and bias states $\{m_i\}$. 

We next turn to the computation of the generating function of $Q$. Multiplying Eq.~\eqref{PQ} with $e^{-p Q}$ and summing over $Q$ yields
\begin{equation}\label{lpbasic}
 \sum_{Q=0}^{\infty} e^{-pQ} P(Q,t,\{x_i\},\{m_i\}) = \langle e^{-pQ}\rangle_{\{x_i\},\{m_i\}} = \left \langle \exp[{-p \sum_{i=1}^N {\cal I}_i(t)}]\right \rangle_{\{x_i\},\{m_i\}} .
 \end{equation}
Since we focus on a non-interacting process, the motion of each particle can be considered independently. We have the identity $e^{-p {\cal I}_i} = 1 -(1-e^{-p}){\cal I}_i$ because the indicator variable ${\cal I}_i$ can only take values $0$ or $1$. Inserting this identity in Eq.~(\ref{lpbasic}) and considering the non-interacting nature of particle motion yield
 \begin{equation}\label{sq-Ii}
  \langle e^{-pQ}\rangle_{\{x_i\},\{m_i\}} = \prod_{i=1}^N\left[1- (1-e^{-p})\langle {\cal I}_i(t) \rangle_{\{x_i\},\{m_i\}} \right].
 \end{equation}
The average $\langle {\cal I}_i(t) \rangle_{\{x_i\},\{m_i\}}$ is the probability that the $i^{\rm th}$ particle starting from the location $x_i$ in the bias state $m_i$ is present on the right side of the origin at time $t$. This quantity is connected to the Green's function $G^{m_i}(x,x_i,t)$ as
\begin{equation}\label{I-propagator}
  \langle {\cal I}_i(t) \rangle_{\{x_i\},\{m_i\}} = \int_0^{\infty}dx~ G^{m_i}(x,x_i,t)  = U^{m_i}(-x_i,t) , \quad x_i < 0 .
 \end{equation}
After inserting Eq.~(\ref{I-propagator}) into Eq.~(\ref{sq-Ii}), we obtain
\begin{equation}\label{history_av}
  \langle e^{-pQ}\rangle_{\{x_i\},\{m_i\}}= \prod_{i=1}^N\left[1- (1-e^{-p})U^{m_i}(-x_i,t)\right]  , \quad x_i < 0.
 \end{equation}
As specified before,~$\overline{\cdots}$ denotes an average over the initial positions and $\overbrace{\cdots}$ denotes an average over the initial bias states of the particles. In the following subsections, we consider the effect of these averages separately. By expanding the generating function in $p$ for each of the four cases, we demonstrate below that the mean of $Q$ remains the same across all initial conditions. However, the variance displays distinct behaviors depending on the initial conditions. 
\subsection{Case 1: Annealed density and annealed magnetization initial conditions}
We first consider an initial condition where the initial positions and bias states of particles are allowed to fluctuate. The flux distribution for this case is denoted as $P_{a,a}(Q,t)$. The moment-generating function for this distribution is given in Eq.~\eqref{a_a_av}.
The position of each particle is distributed uniformly in the box $[-L,0]$. While we initially consider a system of finite size, we eventually take the limit $L \rightarrow \infty$, $N \rightarrow \infty$ with $N/L \rightarrow \rho$ fixed in our analytical calculations.
After averaging over the initial positions in Eq.~\eqref{history_av}, we obtain
\begin{eqnarray}
\overline{\langle e^{-p Q}\rangle_{\{x_i\},\{m_i\}}} &=& \prod_{i=1}^N\left[1- (1-e^{-p}) \overline{U^{m_i}(-x_i,t)}\right]\nonumber\\
&=&\prod_{i=1}^N \left[1- (1-e^{-p}) \int_{-L}^{0}\frac{dx_i}{L}~ U^{m_i}(-x_i,t)  \right] \nonumber\\
&=& \left[1- \frac{1}{L}(1-e^{-p}) \int_0^L dz~U^{m_z}(z,t)  \right]^N ,~z=-x_i.
\label{history_pos_av}
\end{eqnarray}
In this context, $m_z$ represents the bias state of the particle located at $x_i=-z$ at time $t=0$.
Subsequently, after averaging over the initial bias states in the above equation, we obtain
\begin{equation}
\overbrace{ \overline{\langle e^{-p Q}\rangle_{\{x_i\},\{m_i\}}}} = \left[1- \frac{1}{L}(1-e^{-p}) \int_0^L dz~\left(f^+U^{+}(z,t)+f^-U^{-}(z,t)\right)  \right]^N .
\end{equation}
In the limit as $N$ and $L$ tend to infinity while keeping the ratio $\rho = N/L$ fixed, we obtain
\begin{eqnarray}
\sum_{Q=0}^{\infty} e^{-pQ} P_{a,a}(Q,t) &=& \overbrace{ \overline{\langle e^{-p Q}\rangle_{\{x_i\},\{m_i\}}}} \rightarrow \exp\left[-\mu(t) ~ (1-e^{-p})\right],
\label{a_a_gf}
\end{eqnarray}
where
\begin{equation}
  \mu(t)=\rho \int_0^{\infty} dz ~ \left[ f^+U^{+}(z,t)+f^-U^{-}(z,t)\right].
  \label{mu_t}
\end{equation}
This corresponds precisely to the moment-generating function of a Poisson distribution. Consequently, $P_{a,a}(Q=N,t)$ always follows a Poisson distribution with
\begin{equation}
P_{a,a}(Q=N,t) = e^{-\mu(t)} \frac{\mu(t)^N}{N!},~ N~=0,1,2,\cdots~.
\label{Paa}
\end{equation}
The expressions for the mean and variance are thus given as
\begin{eqnarray}
&&\langle Q \rangle_{a,a} = \mu(t),  \nonumber\\
&&\sigma_{a,a}^2 = \langle Q^2 \rangle_{a,a} - \langle Q \rangle^2_{a,a} = \mu(t), 
\end{eqnarray}
where $\mu(t)$ is defined in Eq.~\eqref{mu_t}. The mean and the variance are the same for annealed density and annealed magnetization initial conditions which has a particularly simple form in Laplace space. Using the expression in Eq.~\eqref{Uplus_minus_int} to compute the Laplace transform of Eq.~\eqref{mu_t}, we obtain
\begin{eqnarray}
 \tilde \mu(s)=\rho \left[f^+ \frac{ v \left(\sqrt{s (s+2
   \gamma )}+s\right)}{2 s^2
   (s+2 \gamma )}+f^- \frac{v \left(\sqrt{s (s+2
   \gamma )}-s\right)}{2 s^2
   (s+2 \gamma )} \right].
  \label{mu_s}
\end{eqnarray}
This expression can be inverted exactly yielding the expression for the mean and variance as in Eq.~\eqref{mu_t_exact}. 
\subsection{Case 2: Annealed density and quenched magnetization initial conditions}
We next consider an initial condition where the initial positions of the particles are allowed to fluctuate, but the bias states are fixed. The flux distribution for this case is represented as $P_{a,q}(Q,t)$. The corresponding moment-generating function is given in Eq.~\eqref{a_q_av}. Substituting Eq.~\eqref{history_pos_av} in Eq.~\eqref{a_q_av}, we obtain
\begin{align}
\exp \left[ \overbrace{\ln \overline{ \langle e^{-p Q}\rangle_{\{x_i\},\{m_i\}}}}\right]&=\exp[\overbrace{N \ln \left[1- \frac{1}{L}(1-e^{-p}) \int_0^L dz~U^{m_z}(z,t)  \right] }]\nonumber\\
&= \exp[N f^+\ln \left[1- \frac{1}{L}(1-e^{-p}) \int_0^L dz~U^+(z,t) \right]]\times\nonumber\\
&\exp[ N f^-\ln \left[1- \frac{1}{L}(1-e^{-p}) \int_0^L dz~U^-(z,t)  \right]]\nonumber\\
&\rightarrow \exp[-\rho  (1-e^{-p}) \int_0^\infty dz~\left[f^+U^+(z,t)+f^-U^-(z,t)\right]]  
\nonumber\\
&=\exp[-\mu(t)(1-e^{-p})],
\end{align}
which is same as Eq.~\eqref{a_a_gf} and $\mu(t)$ is defined in Eq.~\eqref{mu_t}. Consequently, $P_{a,q}(Q=N,t)$ is also a Poisson distribution with
\begin{equation}
P_{a,q}(Q=N,t) = e^{-\mu(t)} \frac{\mu(t)^N}{N!},~ N~=0,1,2,\cdots~.
\label{Paq}
\end{equation}

In the large system size limit ($L \rightarrow \infty,~N \rightarrow \infty,~N/L \rightarrow \rho=\text{fixed}$), the distribution $P_{a,q}(Q,t)$ is equivalent to the distribution $P_{a,a}(Q,t)$. The distinction between keeping the velocities constant or allowing them to fluctuate does not matter in the annealed density setting, as the initial positions of particles are randomized. Thus we obtain the expressions for the mean and variance as
\begin{eqnarray}
\langle Q \rangle_{a,q} = \mu(t),  \\
\sigma_{a,q}^2 = \langle Q^2 \rangle_{a,q} - \langle Q \rangle^2_{a,q}  =\mu(t), 
\end{eqnarray}
where the expression for $\mu(t)$ is provided in Eq.~\eqref{mu_t_exact}. 
\subsection{Case 3: Quenched density and quenched magnetization initial conditions}
We next consider an initial condition where the initial positions and bias states of particles are fixed. The flux distribution for this case is denoted as $P_{q,q}(Q,t)$. The moment-generating function for this flux distribution is given as in Eq.~\eqref{q_q_qv}. By taking the logarithm of both sides of Eq.~\eqref{history_av}, we obtain
\begin{equation}\label{Nlog-quenched}
{\ln}\left[\langle e^{-pQ}\rangle_{\{x_i\},\{m_i\}} \right] = \sum_{i=1}^N {\ln}\left[1-(1-e^{-p})U^{m_i}(-x_i,t) \right].
\end{equation}
Subsequently, we compute the average over the initial positions. We independently and uniformly select each $x_i$ from the interval $[-L, 0]$ and then take the limit as $N \to \infty$, $L \to \infty$, while keeping the ratio $\rho = N/L$ constant. This yields
\begin{eqnarray}\label{Nlog-qu-avg}
\overline{\ln\left[\langle e^{-pQ}\rangle_{\{x_i\},\{m_i\}} \right]}&=& \frac{N}{L}\int_{-L}^0 dx_i ~{\ln}\left[1-(1-e^{-p})U^{m_i}(-x_i,t) \right] \nonumber\\ &\rightarrow& \rho \int_0^\infty dz \ln \left[ 1 - (1-e^{-p}) U^{m_z}(z,t)\right] ,~z=-x_i.
\end{eqnarray} 
Next performing the average over initial bias states, we obtain
\begin{eqnarray}
\overbrace{\overline{{\ln}\left[\langle e^{-pQ}\rangle_{\{x_i\},\{m_i\}} \right]}}&=&  \rho f^+ \int_0^\infty dz \ln \left[ 1 - (1-e^{-p}) U^+(z,t)\right] \nonumber\\&+&\rho f^- \int_0^\infty dz \ln \left[ 1 - (1-e^{-p}) U^-(z,t)\right] .
\end{eqnarray} 
The above expression represents the cumulant-generating function for the distribution $P_{q,q}(Q,t)$. To obtain the cumulants, we collect terms that occur at the same powers of $p$. This allows us to derive the expressions for the mean and variance, which are given as
\begin{align}
\langle Q \rangle_{q,q}  &=\mu(t), \nonumber \\
\sigma_{q,q}^2 &= \langle Q^2 \rangle_{q,q} - \langle Q \rangle^2_{q,q} \nonumber\\
&= \rho \int_0^\infty dz ~\left[f^+ U^+(z,t)(1-U^+(z,t)) \right] + \rho \int_0^\infty dz~\left[f^- U^-(z,t)(1-U^-(z,t)) \right]  ,\nonumber\\
\label{var_q_q}
\end{align}
where the expression for $\mu(t)$ is provided in Eq.~\eqref{mu_t_exact}. To compute the variance, we take a Laplace transform of the expression for variance provided in Eq.~\eqref{var_q_q} which yields
\begin{eqnarray}
\tilde \sigma_{q,q}^2 (s) =\rho \left( f^+ \tilde T_1 (s)+f^-\tilde T_2 (s) \right),
\label{var_q_q_s}
\end{eqnarray}
where 
\begin{equation}
 \tilde T_1 (s)   = \mathcal{L}\left[ \int_0^\infty dz~\left[ U^+(z,t)(1-U^+(z,t)) \right]  \right]=\int_0^\infty dz~\left[\tilde U^+(z,s)-\tilde V^+(z,s)\right],
\end{equation}
and
\begin{equation}
 \tilde T_2 (s)   = \mathcal{L}\left[ \int_0^\infty dz~\left[U^-(z,t)(1-U^-(z,t)) \right]   \right]=\int_0^\infty dz~\left[\tilde U^-(z,s)-\tilde V^-(z,s)\right].
\end{equation}
The functions $\tilde V^+(z,s)$ and $\tilde V^-(z,s)$ are defined in Eq.~\eqref{Vplus_minus_def}.
Substituting the expressions provided in Eqs.~\eqref{Uplus_minus_int}~and~\eqref{Vplus_minus_int} in the above equations, we obtain
\begin{eqnarray}
\tilde T_1 (s)  = 
\frac{v}{4 s (s+2 \gamma )}+\frac{v \gamma }{s (s+2 \gamma ) \sqrt{s (s+4 \gamma )}}-\frac{v
   K\left(-\frac{8 \gamma  (s+2 \gamma )}{s^2}\right)}{2 \pi  s (s+2 \gamma )},
   \label{T1}
\end{eqnarray}
and
\begin{eqnarray}
\tilde T_2(s)  = 
-\frac{v}{4 s (s+2 \gamma )}+\frac{v \gamma }{s (s+2 \gamma ) \sqrt{s (s+4 \gamma )}}+\frac{v
   K\left(-\frac{8 \gamma  (s+2 \gamma )}{s^2}\right)}{2 \pi  s (s+2 \gamma )}.
   \label{T2}
\end{eqnarray}
Combining Eqs.~\eqref{var_q_q_s},~\eqref{T1}~and~\eqref{T2}, we obtain the expression for the variance in Laplace space as in Eq.~\eqref{sigma_q_q}. Performing series expansions, we obtain the following expressions in the small and large $s$ limits,
\begin{eqnarray} 
\tilde \sigma_{q,q}^2 (s)&\xrightarrow[s \rightarrow 0]{}&\frac{\rho v}{4 \sqrt{\gamma } s^{3/2}},\nonumber\\
\tilde \sigma_{q,q}^2 (s)&\xrightarrow[s \rightarrow \infty]{}&\left(3 f^++f^- \right) \frac{\rho v \gamma }{2 s^3}.
\end{eqnarray}
The small and large $s$ limits correspond to the large and small time behaviors respectively. Upon inversion of the above expressions, we obtain the limiting behaviors listed in Table~\ref{table1}.

In particular, for the symmetric case where $f^+=f^-=1/2$, the variance assumes a very simple form in Laplace space. Substituting $f^+=f^-=1/2$ in Eq.~\eqref{sigma_q_q} yields
\begin{eqnarray}
\tilde \sigma_{q,q}^2 (s) =\rho \frac{v \gamma }{s^{3/2} (s+2 \gamma ) \sqrt{ (s+4 \gamma )}} .
\label{sigma_q_q_sym_laplace}
\end{eqnarray}
In order to obtain the behavior in time, we perform the inverse Laplace transform of the above expression.
It is convenient to break up the expression as
\begin{eqnarray}
\tilde \sigma_{q,q}^2 (s) = \tilde{f}(s) . \tilde{g}(s),
\end{eqnarray}
with
\begin{eqnarray}
\nonumber
\tilde{f}(s) &=& \rho \frac{ v\gamma}{ s^{3/2}\sqrt{s+4 \gamma }} ,\\
\tilde{g}(s) &=& \frac{1}{s+2 \gamma}.
\end{eqnarray}
Each of these expressions can be inverted individually yielding
\begin{eqnarray}
\nonumber
f(t) &=& \mathcal{L}^{-1}[\tilde{f}(s)] = \rho  v \gamma e^{-2 \gamma  t} t\left[ \pmb{I}_0(2 t \gamma )+ \pmb{I}_1(2 t \gamma )\right],\\
g(t) &=&\mathcal{L}^{-1}[\tilde{g}(s)] =  e^{-2 \gamma  t}.
\end{eqnarray}
Using the convolution theorem
\begin{eqnarray}
\mathcal{L}^{-1}[\tilde{f}(s) \tilde{g}(s)] = \int_{0}^{t}d\tau~f(\tau)g(t-\tau),
\end{eqnarray}
we arrive at the following expression for the variance for the quenched case
\begin{eqnarray}
\sigma_{q,q}^2 (t) = \rho  v \gamma  e^{-2 \gamma t} \int_{0}^{t} d\tau~  \tau \left[ \pmb{I}_0(2 \tau \gamma )+\pmb{I}_1(2 \tau \gamma )\right].
\end{eqnarray}
Performing this integral, we arrive at the exact expression provided in Eq.~\eqref{sigma_q_q_sym}.
\subsection{Case 4: Quenched density and annealed magnetization initial conditions}
Finally, we consider an initial condition where the initial positions of the particles are fixed, but the bias states are allowed to fluctuate. The flux distribution for this case is denoted as $P_{q,a}(Q,t)$. The moment-generating function associated with this process is given in Eq.~\eqref{q_a_av}.
After calculating the average across initial bias states in Eq.~\eqref{history_av}, we arrive at 
\begin{align}
\overbrace{\langle e^{-p Q}\rangle_{\{x_i\},\{m_i\}}} &= \prod_{i=1}^N\left[1- (1-e^{-p}) \overbrace{U^{m_i}(-x_i,t)}\right]\nonumber\\&=\prod_{i=1}^N\left[1- (1-e^{-p}) \left(f^+U^+(-x_i,t)+f^-U^-(-x_i,t)\right)\right].
\end{align}
Taking a logarithm of the moment generating function yields the cumulant generating function. From the above expression, we thus directly compute the cumulant generating function as
\begin{align}
\overline{\ln \overbrace{ \langle e^{-p Q}\rangle_{\{x_i\},\{m_i\}}}}&= \frac{N}{L}\int_{-L}^0 dx_i ~{\ln}\left[1-(1-e^{-p})\left(f^+U^+(-x_i,t)+f^-U^-(-x_i,t)\right) \right] \nonumber\\
&\rightarrow \rho \int_0^\infty dz \ln \left[ 1 - (1-e^{-p}) \left(f^+U^+(z,t)+f^-U^-(z,t)\right)\right] ,~z=-x_i.\nonumber\\
\end{align}
Collecting the terms that appear in the first and second powers of $p$, we derive the expressions for the mean and variance of $Q$ as
\begin{align}
\langle Q \rangle_{q,a} &=\mu(t),  
\nonumber\\
\sigma_{q,a}^2 &= \langle Q^2 \rangle_{q,a} - \langle Q \rangle^2_{q,a} \nonumber\\
&= \rho \int_0^\infty dz~\left[ \left( f^+U^{+}(z,t)+f^-U^{-}(z,t)\right)\left(1-\left( f^+U^{+}(z,t)+f^-U^{-}(z,t)\right)\right)\right]  , 
\end{align}
where the expression for $\mu(t)$ is provided in Eq.~\eqref{mu_t_exact}. Since the computation of the variance in real space is difficult, we turn to the computation of the variance in Laplace space. From the above expression, we obtain
\begin{align}
\tilde \sigma_{q,a}^2 (s) &=\rho \int_0^\infty dz~\Big[  f^+\tilde U^{+}(z,s)+f^-\tilde U^{-}(z,s)-{f^+}^2 \tilde V^{+}(z,s)- {f^-}^2\tilde V^{-}(z,s)\nonumber\\
&-2 f^+f^-V^{\text{cross}}(z,s)\Big] .
\label{var_q_a_laplace_exp}
\end{align}
The integral of each term in the above expression are provided in Eqs.~\eqref{Uplus_minus_int},~\eqref{Vplus_minus_int}~and~\eqref{Vcross_int}. Combining these results, we obtain the exact expression for the variance in Laplace space as in Eq.~\eqref{sigma_q_a}.  The asymptotic forms of this expression are remarkably simple. These are given as
\begin{eqnarray} 
\tilde \sigma_{q,a}^2 (s)&\xrightarrow[s \rightarrow 0]{}&\frac{\rho v}{4 \sqrt{\gamma } s^{3/2}},\nonumber\\
\tilde \sigma_{q,a}^2 (s)&\xrightarrow[s \rightarrow \infty]{}&f^+f^-\frac{ \rho v}{s^2}++\frac{ \rho  v \gamma }{2s^3} \left(3 f^+-f^-\right) \left(f^+-f^-\right) .
\end{eqnarray}
The inverse Laplace transforms of the above expressions yield the limiting behaviors in time as listed in Table~\ref{table1}. Interestingly, we observe that the short-time behavior of current fluctuations is determined by the product of the initially present fraction of particles in the positive and negative states. This sets it apart from other cases where the leading linear behavior of current fluctuations at short times is solely determined by the fraction of positive particles. Therefore this specific initial condition induces cross-correlations between positive and negative states, as also evidenced by the presence of the cross-term $V^{\text{cross}}$ in Eq.~\eqref{var_q_a_laplace_exp}. 

For the symmetric case where $f^+=f^-=1/2$, the expression in Eq.~\eqref{sigma_q_a} reduces to
\begin{eqnarray}
\tilde \sigma_{q,a}^2 (s) = \rho \frac{ v}{2 s^{3/2}} \left(\frac{\sqrt{s+4 \gamma }}{2( s+2\gamma)}\right).
\label{eq_exact_quenched_laplace}
\end{eqnarray}
We can rewrite this expression as
\begin{eqnarray}
\tilde \sigma_{q,a}^2 (s) = 
 \rho \frac{v}{2 s^{3/2}}\frac{1}{\sqrt{(s+2 \gamma)}}\left(\frac{1}{2}\frac{\sqrt{(s+4 \gamma)}}{\sqrt{(s+2 \gamma)}}\right).
\end{eqnarray}
We should compare this with the exact expression for the variance for the annealed case provided in Eq.~\eqref{mu_s} with $f^+=f^-=1/2$.
We, therefore, have the identity
\begin{eqnarray}
\frac{\tilde \sigma_{q,a}^2 (s)}{ { \tilde \sigma_{a,a}^2 (s)}} = 
\left(\frac{1}{2}\frac{\sqrt{(s+4 \gamma)}}{\sqrt{(s+2 \gamma)}}\right).
\end{eqnarray}
In the large $s$ limit, this yields a factor of $2$, and in the small $s$ limit this yields a factor of $\sqrt{2}$ as also observed in previous studies~\cite{banerjee2020current,jose2023generalized}.

In order to obtain the behavior in time, we perform the inverse Laplace transform of the expression in Eq.~\eqref{eq_exact_quenched_laplace}.
It is convenient to break up the expression as
\begin{eqnarray}
\tilde \sigma_{q,a}^2 (s) &=& \tilde{h}(s) . \tilde{q}(s),
\end{eqnarray}
with
\begin{eqnarray}
\nonumber
\tilde{h}(s) &=& \rho \frac{ v}{2 s^{3/2}} \left(\sqrt{s+4 \gamma }\right),\\
\tilde{q}(s) &=& \frac{1}{2(s+2 \gamma)}.
\end{eqnarray}
Each of these expressions can be inverted individually yielding
\begin{eqnarray}
\nonumber
h(t) &=& \mathcal{L}^{-1}[\tilde{h}(s)] = \frac{\rho  v}{2} e^{-2 \gamma  t} \left[(4 \gamma  t+1) \pmb{I}_0(2 t \gamma )+4
   \gamma  t \pmb{I}_1(2 t \gamma )\right],\\
q(t) &=&\mathcal{L}^{-1}[\tilde{q}(s)] = \frac{1}{2} e^{-2 \gamma  t}.
\end{eqnarray}
Using the convolution theorem
\begin{eqnarray}
\mathcal{L}^{-1}[\tilde{h}(s) \tilde{q}(s)] &=& \int_{0}^{t}d\tau~h(\tau)q(t-\tau),
\end{eqnarray}
we arrive at the following expression for the variance for the quenched case
\begin{eqnarray}
\sigma_{q,a}^2 (t) &=& \frac{\rho  v }{4} e^{-2 \gamma t} \int_{0}^{t} d\tau~  \left[(4 \gamma  \tau+1) \pmb{I}_0(2 \tau \gamma )+4
   \gamma  \tau \pmb{I}_1(2 \tau \gamma )\right].
\end{eqnarray}
Performing this integral, we arrive at the exact expression in Eq.~\eqref{sigma_q_a_sym}.

\section{Conclusion and discussion}
\label{sec:conclusion}

In this paper, we have studied the fluctuations (variance) of the integrated current $Q$ across the origin up to time $t$ in a one-dimensional system of non-interacting run and tumble particles. Our analysis involved performing annealed and quenched averages over general step initial conditions for both density and magnetization fields associated with the motion of particles. Our analytical findings provide valuable insights into the dynamic behavior of the fluctuations of $Q$. At large times, we observed that these fluctuations grow as $\sqrt{t}$, which indicates the effective diffusive nature of the system during these times. The behavior of the fluctuations at large times is independent of the magnetization initial conditions and depends only on the density initial conditions. Annealed density conditions display larger fluctuations with a factor of $\sqrt{2}$ consistent with previous findings in the literature~\cite{derrida2009current2,krapivsky2012fluctuations,banerjee2020current,banerjee2022role}.

The situation is significantly different at short times, where the magnetization initial conditions play a crucial role in determining the growth exponent of the fluctuations. When we have a non-zero fraction $f^+$ of particles in the $+$ velocity state at time $t=0$, the fluctuations display a quadratic $t^2$ growth for quenched density and quenched magnetization initial conditions. On the other hand, if we employ annealed initial conditions in either of the fields, the fluctuations exhibit a linear $t$ growth. Notably, the prefactor for each of these cases strongly depends on the type of density and magnetization initial conditions used. Interestingly, when $f^+=0$, meaning there are no particles in the $+$ velocity state at time $t=0$, the fluctuations exhibit a $t^2$ growth regardless of the type of initial conditions employed. 

Our results highlight how slight variations in initial conditions can result in significant disparities in the behavior of active systems over time. Although the techniques outlined in this paper are specific to non-interacting RTPs in one dimension, they can also be extended to systems with multiple degrees of freedom at the particle level, even in higher spatial dimensions. One approach to defining a current in higher dimensions involves considering $N$ particles uniformly distributed within a specific region $\mathcal{R}$ in space. We define the space outside this region as $\mathcal{S}$. At any given time $t$, the current of the system, denoted as $Q$, represents the number of particles that have exited region $\mathcal{R}$ up to time $t$, or equivalently, the number of particles present in the region $\mathcal{S}$ at time $t$. It is possible to generalize many of our results to such a situation, which could help in understanding the transport properties of non-interacting particles and the influence of various initial conditions and geometries. Furthermore, despite the effective diffusive behavior of the fluctuations at late times, previous studies~\cite{banerjee2020current} have shown that the fingerprints of activity are visible in the full large deviation function in the quenched density and the annealed magnetization setting. It would therefore be interesting to study the large deviation function for the case where both the density and magnetization fields are quenched and analyze how the effects of activity persist in such cases.

\ack
We thank S.~N. Majumdar, G. Schehr, and R. Maharana for useful discussions. The work of K.~R. was partially supported by the SERB-MATRICS grant MTR/2022/000966. This project was funded by intramural funds at TIFR Hyderabad from the Department of Atomic Energy (DAE), Government of India.

\appendix

\section{Laplace transform of the square of a function }
\label{appendix_laplace_transforms_sq}
In this Appendix, we show that the knowledge of the Laplace transform of a function $U(z,t)$ also yields the Laplace transform of the square of the function. Let us define
\begin{equation}
    \tilde V(z,s)=\mathcal{L}\left[U(z,t)^2 \right],
\end{equation}
as the Laplace transform of the square of the function $U(z,t)$. The expression for $V(z,s)$ can be rewritten as
\begin{eqnarray}
\tilde V(z,s)&=&\int_{0}^{\infty}\int_{0}^{\infty}dt~ dt'~U(z,t)U(z,t')e^{-\frac{s}{2}(t+t')}\delta(t-t').
\end{eqnarray}
Using the integral representation of the Dirac delta function in the above expression, we obtain
\begin{align}
\tilde V(z,s)&= \frac{1}{2 \pi}\int_{0}^{\infty}\int_{0}^{\infty}\int_{-\infty}^{\infty}dt~dt'~dk~U(z,t)U(z,t')e^{-\frac{s}{2}(t+t')}e^{i k (t- t')}\nonumber\\
&=\frac{1}{2 \pi}\int_{-\infty}^{\infty} dk~ \Biggl(\int_{0}^{\infty}dt~U(z,t)e^{-(\frac{s}{2}-i k)t}\Biggr)\Biggl(\int_{0}^{\infty}dt'~U(z,t')e^{-(\frac{s}{2}+i k)t'}\Biggr).
\end{align}
This is the product of two Laplace transforms, which can be written as
\begin{eqnarray}
\tilde V(z,s)
=\frac{1}{2 \pi}\int_{-\infty}^{\infty}dk~ \tilde U\left(z,\frac{s}{2}-i k\right)  \tilde U\left(z,\frac{s}{2}+i k\right).
\label{final_uzs}
\end{eqnarray}
The above expression is extremely useful as it directly computes the Laplace transform of the square of a function from the knowledge of the Laplace transform of the function itself.
\section{Details of calculations}
\label{details_calc}
In this Appendix, we provide details regarding the calculations presented in the main text based on the identity provided in Eq.~\eqref{final_uzs}.

Let us first derive the expression provided in Eq.~\eqref{Vs_int}.
Using Eqs.~\eqref{uzs} and~\eqref{final_uzs}, the explicit expression of $\tilde V^0(z,s)$ can be computed as
\begin{align}
\tilde V^0(z,s)&=\frac{1}{2 \pi}\int_{-\infty}^{\infty}dk~ \frac{\exp \left(-\frac{z \sqrt{\left(\frac{s}{2}-i k\right) \left(\frac{s}{2}-i k+2
   \gamma \right)}}{v}-\frac{z \sqrt{\left(\frac{s}{2}+i k\right) \left(\frac{s}{2}+i k+2
   \gamma \right)}}{v}\right)}{4 \left(\frac{s}{2}-i k\right) \left(\frac{s}{2}+i k\right)}.
\end{align}
We next compute the integral of the above function over $z$. This yields
\begin{align}
\int_0^\infty dz~ \tilde V^0(z,s) 
 =  \frac{1}{2 \pi} \int_0^\infty  dz \int_{-\infty}^{\infty}dk~ \frac{\exp \left(-\frac{z \sqrt{\left(\frac{s}{2}-i k\right) \left(\frac{s}{2}-i k+2
   \gamma \right)}}{v}-\frac{z \sqrt{\left(\frac{s}{2}+i k\right) \left(\frac{s}{2}+i k+2
   \gamma \right)}}{v}\right)}{4 \left(\frac{s}{2}-i k\right) \left(\frac{s}{2}+i k\right)}.
\end{align}
The integral becomes simpler if we first perform the $z$ integral and this yields
\begin{small}
\begin{align}
\int_0^\infty dz~\tilde V^0(z,s) 
 =
 \frac{1}{2 \pi} \int_{-\infty}^{\infty}dk~ \frac{2 v}{\left(s^2+4 k^2\right) \left(\sqrt{(s- 2i k) (s+4 \gamma
   -2ik)}+\sqrt{(s+2i k) (s+4 \gamma +2ik)}\right)}\label{sigma_qu_rtp}
\end{align}
\end{small}
The integral above can be done in closed form and has a particularly simple answer. Let us define the function $I^0(k)$ as
\begin{align}
I^0(k)=
 \int dk~ \frac{2 }{\left(s^2+4 k^2\right) \left(\sqrt{(s- 2i k) (s+4 \gamma
   -2ik)}+\sqrt{(s+2i k) (s+4 \gamma +2ik)}\right)}.
\end{align}
\begin{figure} [t!]
\centering
 \includegraphics[width=0.6\linewidth]{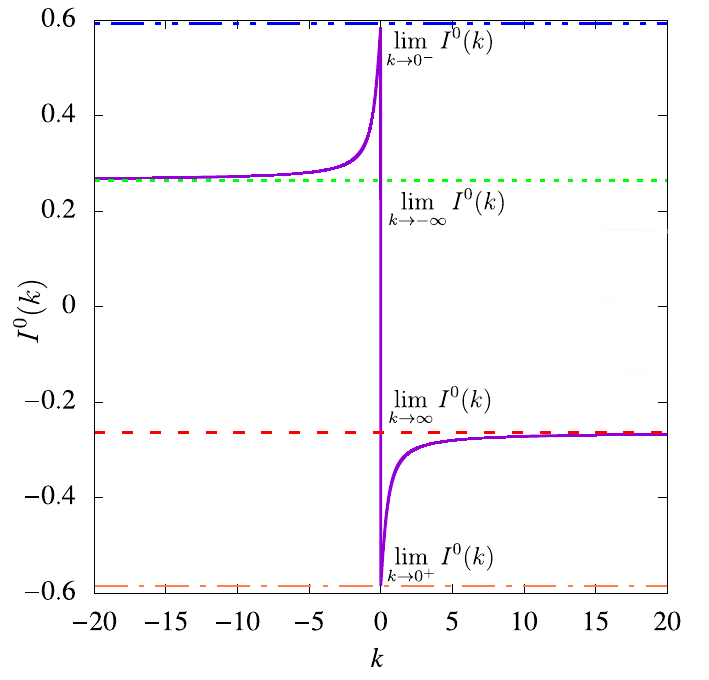}
\caption{The integral $I^0(k)$ provided in Eq.~\eqref{indefinite_integral_rtp} plotted as a function of $k$ for fixed $s=1,~\gamma=1$~(solid purple curve). The integral has a discontinuity across the origin. The limiting forms $\lim_{k \rightarrow \infty} I^0(k),~lim_{k \rightarrow -\infty} I^0(k),~lim_{k \rightarrow 0^+} I^0(k)$ and ~$lim_{k \rightarrow 0^-} I^0(k)$ are provided in Eqs.~\eqref{limits_infty}~and~\eqref{limits_0}.}
\label{fig:i(k)}
\end{figure}
The indefinite integral $I^0(k)$ can be explicitly computed as
\begin{small}
\begin{align}
I^0(k) &=\frac{1}{2 s^{3/2} (s+2 \gamma ) \sqrt{-(2
   k+i s)}}\times  \nonumber\\&\Biggl[
   \sqrt{(2 k+i s) (s+2 \gamma )} \left(\tanh
   ^{-1}\left(\frac{\sqrt{(2 k+i s) (s+2 \gamma )}}{\sqrt{s (2 k+i
   (s+4 \gamma ))}}\right)-\tanh ^{-1}\left(\frac{\sqrt{(2 k-i s) (s+2\gamma )}}{\sqrt{s (2 k-i (s+4 \gamma ))}}\right)\right) 
   \nonumber\\
   &+\sqrt{(2k+i s) (s+4 \gamma )} \left(\tanh ^{-1}\left(\frac{\sqrt{(2 k-i s)
   (s+4 \gamma )}}{\sqrt{s (2 k-i (s+4 \gamma ))}}\right)-\tanh
   ^{-1}\left(\frac{\sqrt{(2 k+i s) (s+4 \gamma )}}{\sqrt{s (2 k+i
   (s+4 \gamma ))}}\right)\right)\Biggr].
\label{indefinite_integral_rtp}
\end{align}
\end{small}
The function $I^0(k)$ has the typical behavior provided in Fig.~\ref{fig:i(k)}. The definite integral in Eq.~\eqref{sigma_qu_rtp} can be computed simply as
\begin{align}
\int_0^\infty dz~\tilde V^0(z,s)&=
   \frac{ v}{2 \pi} \Bigl(\lim_{k \rightarrow \infty} I^0(k)-\lim_{k \rightarrow 0^+} I^0(k)+\lim_{k \rightarrow 0^-} I^0(k)-\lim_{k \rightarrow -\infty} I^0(k)\Bigr).  
   \label{vs_ik}
\end{align}
We need to extract the asymptotic behaviors of $I^0(k)$. These can be computed as
\begin{equation} 
I^0(k)\xrightarrow[k \rightarrow \pm\infty]{}\ \mp \frac{\pi  \left(\sqrt{\frac{s+4 \gamma }{s+2 \gamma }}-1\right)}{2
   s^{3/2} \sqrt{s+2 \gamma }},
\label{limits_infty}
\end{equation}
and
\begin{equation} 
I^0(k)\xrightarrow[k \rightarrow 0^{\pm}]{}\ \mp \frac{\pi  \sqrt{\frac{s+4 \gamma }{s+2 \gamma }}}{4 s^{3/2} \sqrt{s+2
   \gamma }}.
\label{limits_0}
\end{equation}
Combining Eqs.~\eqref{vs_ik},~\eqref{limits_infty}~and~\eqref{limits_0}, we obtain the desired result in Eq.~\eqref{Vs_int}.

Let us next derive the expression provided in Eq.~\eqref{Vplus_minus_int}. One can use the Eqs.~\eqref{uzs_plus_minus} and~\eqref{final_uzs} and proceed exactly like the symmetric case. It is possible to show that
\begin{align}
\int_0^\infty dz~\tilde V^\pm(z,s)&=
   \frac{ v}{2 \pi} \Bigl(\lim_{k \rightarrow \infty} I^\pm(k)-\lim_{k \rightarrow 0^+} I^\pm(k)+\lim_{k \rightarrow 0^-} I^\pm(k)-\lim_{k \rightarrow -\infty} I^\pm(k)\Bigr), 
   \label{vplus_ik}
\end{align}
where $I^\pm(k)$ has the limiting behaviors
\begin{equation} 
I^+(k)\xrightarrow[k \rightarrow \pm\infty]{}\ \mp \left(\frac{2 \pi  \gamma }{s^{3/2} (s+2 \gamma )
   \sqrt{s+4 \gamma }}- \frac{\pi +2 K\left(-\frac{8
   \gamma  (s+2 \gamma )}{s^2}\right)}{4 s (s+2
   \gamma )}\right),
\label{limits_infty_plus}
\end{equation}
\begin{equation} 
I^-(k)\xrightarrow[k \rightarrow \pm\infty]{}\ \mp \left(\frac{2 \pi  \gamma }{s^{3/2} (s+2 \gamma )
   \sqrt{s+4 \gamma }}+\frac{\pi +2 K\left(-\frac{8
   \gamma  (s+2 \gamma )}{s^2}\right)}{4 s (s+2
   \gamma )}\right),
\label{limits_infty_minus}
\end{equation}

\begin{equation} 
I^+(k)\xrightarrow[k \rightarrow 0^{\pm}]{}\ \mp \frac{\pi  \left(\frac{1}{2 \sqrt{s+2 \gamma
   }}+\frac{\gamma }{(s+2 \gamma ) \sqrt{s+4 \gamma
   }}\right)}{s^{3/2}},
\label{limits_0_plus}
\end{equation}
and
\begin{equation} 
I^-(k)\xrightarrow[k \rightarrow 0^{\pm}]{}\ \mp \frac{\pi  \left(\frac{1}{2 \sqrt{s+2 \gamma
   }}+\frac{\gamma }{(s+2 \gamma ) \sqrt{s+4 \gamma
   }}\right)}{s^{3/2}}.
\label{limits_0_minus}
\end{equation}
Combining Eqs.~\eqref{vplus_ik}-\eqref{limits_0_minus}, we obtain the result provided in Eq.~\eqref{Vplus_minus_int}.

\section*{References}
\bibliographystyle{unsrt}
\bibliography{bibtex.bib}

\end{document}